\documentstyle[aps,twocolumn,prl]{revtex} 

\begin{document}
\draft

\twocolumn[\hsize\textwidth\columnwidth\hsize\csname 
@twocolumnfalse\endcsname 

\title{Electronic correlation effects and the Coulomb gap at finite temperature}

\author{B.~Sandow${^1}$, K.~Gloos$^{2, 3}$, R.~Rentzsch${^1}$, A.~N.~Ionov${^4}$, and W.~Schirmacher$^{5}$}

\address{$^1$ Institut f\"ur Experimentalphysik, Freie Universit\"at Berlin, D-14195 Berlin, Germany}
\address{$^2$ Institut f\"ur Festk\"orperphysik, Technische Universit\"at Darmstadt, D-64289 Darmstadt, Germany}
\address{$^3$ Department of Physics, University of Jyv\"askyl\"a, FIN-40351 Jyv\"askyl\"a, Finland}
\address{$^4$ A.~F.~Ioffe Physico-Technical Institute, Russian Academy of Sciences, St. Petersburg, Russia}
\address{$^5$ Physik-Dept.~E13, Technische Universit\"at M\"unchen, D-85747 Garching, Germany}

\date{\today}
\maketitle

\begin{abstract}
We have investigated the effect of the long-range Coulomb interaction
on the one-particle excitation spectrum of n-type Germanium, using
tunneling spectroscopy on mechanically controllable break junctions.
The tunnel conductance was measured as a function of energy and
temperature. At low temperatures, the spectra reveal a minimum at
zero bias voltage due to the Coulomb gap. In the temperature range
above 1\,K the Coulomb gap is filled by thermal excitations. This behavior is
reflected in the temperature dependence of the variable-range hopping
resitivity measured on the same samples: Up to a few degrees Kelvin
the Efros-Shkovskii ln$R \propto T^{-1/2}$ law is obeyed, whereas at
higher temperatures deviations from this law are observed, indicating
a cross-over to Mott's ln$R \propto T^{-1/4}$ law. The mechanism of
this cross-over is different from that considered previously in the
literature.

\end{abstract}


] 
\narrowtext 


The electronic density of states (DOS) near the Fermi level is an
important physical quantity for understanding electrical transport
mechanisms in strongly localized systems\cite{Anderson58}, like
impurity bands in doped semiconductors. They consist of sites with
random positions and random energies \cite{Mott49,Shklovskii84}. The
electron wavefunction is localized with a localization radius $a$
smaller than the average nearest-neighbor distance between
sites which is of order $N^{- 1/3}$ ($N\equiv $ concentration
of impurities). At low temperatures the electrical resistivity of
such systems is governed by variable-range hopping (VRH), which means that the
activation energy for a hopping process decreases continuously with
temperature \cite{Mott49,Shklovskii84}. The role of the inter-site
Coulomb interaction between electrons in the hopping regime was first
addressed by Pollak \cite{Pollak70} and by Srinivasan
\cite{Srinivasa71}. They showed that in localized systems the Coulomb
interaction creates a deep depletion of the one-particle DOS near the
Fermi energy $E_{\rm{F}}$. Efros and Shklovskii (ES) called this
depletion 'Coulomb gap' and showed that the DOS near $E_{\rm{F}}$
varies as $g(E) = g_0 \left( E-E_{\rm{F}} \right) ^ {D-1}$ for
dimensionalities $D= 2$ and $D= 3$, respectively \cite{Efros75,Shklovskii84}.
This leads to a VRH hopping resistance $R \propto \exp{ \left( T_{\rm{ES}}
/ T \right)} ^ {1/2}$ \cite{Efros75,Efros79}, in contrast to Mott's
$R \propto \exp{({T_{\rm{M}} } / {T} )} ^ {1/4}$ law \cite{Mott49}
for which a constant DOS $g(E) = g_0$ is assumed.

A cross-over between these two temperature laws has been expected
theoretically \cite{Shklovskii84} and also observed experimentally
\cite{Ionov83,Shlimak83}. According to the traditional
interpretation of this cross-over the energy range of the
phonon-assisted tunneling (hopping) becomes larger than the width of
the Coulomb Gap $\Delta_{\rm{CG}}$ above the cross-over temperature. In this case the Coulomb gap does
not affect the hopping resistance, thus resulting in Mott's law. However,
another kind of crossover from an ES-type to Mott-type
temperature law may occur, namely the filling of the Coulomb gap by
thermal excitations. Such a filling of the Coulomb gap with
increasing temperature was predicted by Monte-Carlo simulations of a
classical Coulomb glass \cite{Sarvestani95,Grannan93,Shlimak95}. More
recently, it was observed by tunneling sprectroscopy on Si:B samples
near the metal-nonmetal transition. The temperature dependence of the
DOS should, in turn, affect the temperature dependence of the bulk
resistivity. Surprisingly, except for Ref. \cite{Shlimak95}, this
effect has not yet been taken into account in the literature.

In this letter we present evidence that in insulating doped Ge the
second mechanism is verified. We show by comparing the temperature
dependence of the DOS, derived by tunneling spectroscopy, with that
of the VRH resistivity that the observed deviations from the ES law
above 1\,K correspond to the thermal filling of the Coulomb gap and
not to the traditional cross-over mechanism.

The electronic DOS in solids can be dicectly probed by tunneling
spectroscopy and photoelectron spectroscopy, for example. But the
small size of the width of the Coulomb gap $\Delta_{\rm{CG}}$
strongly restricts the useful spectroscopic techniques. At present
only the tunneling spectroscopy has the required energy resolution.
Massey and Lee were the first to directly observe the Coulomb gap in
a doped uncompensated semiconductor Si:B
\cite{Massey95,Massey96,Lee99}. They used planar tunnel junctions
between boron-doped Si samples and a Pb counter-electrode with an
insulating dielectric as barrier. The superconducting quasi-particle
DOS of the lead electrode has been observed, proving quantum
tunneling. Suppressing superconductivity of lead in a magnetic field
of $B = 200\,$mT allowed then to measure the DOS of the Si:B
electrode against the constant DOS of normal lead. However, planar
tunnel junctions are difficult to prepare, especially with germanium.

As alternative method we have proposed recently that it is possible to realize tunneling
across a semiconductor break junction due to the lateral confinement
of small point contacts \cite{Sandow98}. This is a well established technique to investigate
superconductors and metals, both in the regime of direct metallic
contact and tunneling across a vacuum barrier \cite{Muller92}. To ensure a voltage drop
confined to the junction itself the contact resistance is set to a
value larger than about 10\,k$\Omega$ at 1\,K. Our junctions have
lateral dimensions of less than about 100\,nm. The hopping length
amout to about 150\,nm at 1\,K increasing towards lower temperatures.
Charge is transported across the  junction by a single
hopping event at a rather well-defined energy. This justifies our
calling the transport tunneling and to interpret the voltage drop as
an excitation energy. On the other hand, the contact diameter must be
large enough to inhibit the formation of a depletion layer. Such a layer
adds an additional tunneling barrier \cite{Sandow00}, and it may also
vary the local DOS in the contact region. We estimate a $\sim 10\,$nm
lower bound for useful junctions.

Our undoped samples were grown by the Chochralskii method using
highly enriched (up to 93 \% $^{74}$Ge) germanium.
Neutron-transmutation doping (NTD) ensured excellent homogeneity of
$^{75}$As as shallow donors \cite{Shlimak83} . The main nuclear
reaction of Ge with thermal neutrons is
$^{74}{\rm{Ge}}\,(n,\gamma)\,^{75}{\rm{Ge}}\,\rightarrow\,^{75}{\rm{As}}$.
As by-product, a small fraction of $^{75}$Ga is produced as
acceptors. This gives a compensation of $K=N_{\rm{Ga}}/N_{\rm{As}} =
12\,\%$, and sets the Fermi level inside the impurity band. The donor
concentration $N_d = N_{\rm{As}}$ was below, but close to, the
disorder-driven metal-insulator transition of Ge at a critical
impurity concentration of $3.4\cdot 10^{17}\,{\rm{cm}}^{-3}$
\cite{IonovB91}.

For our tunneling experiments the samples were cut into $1\times
1\times 10\,$mm$^3$ slabs with a 0.5\,mm deep groove to define the
break position within the (111) cleaving plane of germanium. The
samples were glued onto a flexible bending beam, electrically isolated
but thermally well coupled to the cold plate. They were broken at low
temperatures in the ultra-high vacuum chamber of the refrigerator.
The contact size could be adjusted {\it in situ} with a micrometer
screw and a piezo tube. For further details of the setup see
Ref.~\cite{Gloos99}. The $dI/dU$ spectra of junctions with small
resistance (less than about 100\, k$\Omega$) were obtained by means
of the standard four-terminal method with current biasing. The
current-voltage characteristics of junctions with high resistance
(larger than about 100\,k$\Omega$) was recorded using the standard
two-terminal method with voltage biasing. In the latter case the bulk
samples contributed at most $5\,\%$ to the total resistance.

All investigated Ge break junctions have rather similar
characteristics. Fig.~\ref{fig-1} shows typical spectra of a sample
with $N_d = 1.26\times 10^{-17}\,{\rm{cm}}^{-3}$ as function of
voltage and temperature. The spectra have a pronounced minimum at low
temperatures. We believe that this anomaly represents the Coulomb
gap. Between 100\,mK and 1\,K the spectra depend only weakly on
temperature. Above about $T=1$ K the Coulomb gap becomes filled by
thermal excitations. It has almost vanished at $T = 6\,$K.

In order to investigate how this temperature dependence of the DOS
affects the hopping resistivity, we have measured the resistance of
the bulk sample using the standard four-terminal technique.
Fig.~\ref{resistance} shows the resistance as function
of $T^{-1/2}$ and $T^{-1/4}$, respectively. At low temperatures ln$R
\propto T ^ {-1/2}$, as expected for the ES law. The resistance
deviates from this behavior at $T^{-1/2}< 0.5\,{\rm{K}}^{-1/2}$ or at
$T > 4\,$K, nearly coincidig with the temperature at which the
Coulomb gap is suppressed according to the tunnel data, see
Fig.~\ref{fig-1}.

We have also considered the traditional scenario of the crossover
between the ES and Mott law in which one assumes that the hopping
energy range becomes broader than Coulomb gap \cite{Shklovskii84}.
>From an analysis similar to that of \cite{Shlimak93} for the traditional
cross-over mechanism we obtained a width
the Coulomb gap of $\Delta_{\rm{CG}} = 0.25\,$meV which is
considerably smaller than the observed $\Delta_{\rm{CG}} =
2.5$\,meV. But the latter value would correspond to a much higher
crossover temperature than observed experimentally.

We turn now to a more detailed discussion of the temperature
dependence of the DOS. To extract the DOS from the spectra we first tried to
remove the energy-dependent part at high voltages. Several
possibilities were tried, with only slight variation of the final
result. According to Ref.~\cite{Lee99}, this high-energy tail can be
roughly described by $g(E) \propto 1+\sqrt{(E-E_{\rm{F}})/\delta}$.
In this model, the parameter $\delta$ represents a correlation
energy, which is almost independent of impurity concentration. For
our experiments $\delta \sim 10\,$meV, a rather large value when
compared to the results for Si:B \cite{Lee99}. Alternatively, a
Schottky-type behaviour was used with $dI/dU \propto \exp{ \left(
U/U_{00} \right) }$. The parameter $U_{00}$ may then represent the
properties of an additional barrier due to the depletion layer. As we do
not know which of the two possibilities is correct,  we normalize the spectra at low temperatures with
respect to that corresponding to the highest temperature. The shape of those
normalized spectra is almost flat outside the Coulomb-gap anomaly.
They are then fitted using
\begin{equation}
g(E,T) \propto \gamma(T) +
\left[ 1-\gamma(T) \right] \frac {|E-E_{\rm{F}}|^s} {(\Delta_{\rm{CG}}(T)/2)^s
+ |E-E_{\rm{F}}|^s}
\label{dos}
\end{equation}
The parameter $\gamma$ describes a 'residual' DOS at the Fermi level
and $\Delta_{\rm{CG}}$ is the width of the Coulomb gap (FWHM). The
DOS derived by ES is recovered when $\gamma = 0$ and $s = 2$
\cite{Shklovskii84,Efros75}.

The experimental DOS of our samples strongly deviates
from the simple square-law derived by ES \cite{Efros75,Shklovskii84},
and which was also found experimentally for Si:B
\cite{Massey95,Massey96,Lee99}.
Taking into account the Fermi distribution and the DOS Eq.~\ref{dos}
on both sides of the junction, our analysis yields for the sample far
from the metal-nonmetal transition $s = 3$. Fig.~\ref{fig-3} shows
how $\Delta_{\rm{CG}}$ and $\gamma$ depend on temperature. Both
saturate at low temperatures.
Analytical
as well as numerical simulations for nonmetallic disordered systems
have predicted several different relationships: power laws $g(E,T=0)
\propto (E-E_F) ^{D-1}$ \cite{Efros75,Baranovskii78} and $g(E,T=0)
\propto (E-E_F) ^{2.7 \pm 0.1}$ \cite{Sarvestani95} as well as an
exponential dependence $g(E,T=0) \propto
\exp[-(\Delta/(E-E_{\rm{F}})) ^{1/2}]$ \cite{Efros76}.
A power law with $s=3$ is consistent with the results of
Ref.~\cite{Sarvestani95}.

The temperature dependence of the zero-bias DOS also reveals a
power law $N(E_F, T) \propto T^x$ with an exponent $x = 0.8$. This
differs from $x=2.7$ derived by \cite{Sarvestani95}, but it agrees
quite well with $x =1$ obtained by the simulations of
Ref.~\cite{Grannan93}.

Because of the experimental DOS, especially its strong temperature
dependence, the observed ES-type behaviour of the bulk resistivity
needs an explanation.
We propose the following scenario to interpret our results: $i)$ At
low temperatures the Coulomb gap ist temperature independent. This
corresponds to the ES regime of the VRH conductivity; $ii)$ the power
law behaviour of the residual DOS $\gamma (T)$ causes the transition
from the ES to the Mott law;
$iii)$ at high temperature the DOS is constant and one
expects a Mott-type VRH conductivity.
The mechanism of the transition between ES- and Mott-type
behaviour is therefore completely different from that considered in
Refs.~\cite{Shklovskii84,Shlimak93}

To summarize, the tunnel conductance of small break junctions of our
germanium samples shows a minimum of the DOS near the Fermi level.
This minimum represents the Coulomb correlation gap. Up to about
1\,K, the width of this anomaly depends only weakly on $T$. This is
the ES-regime. Above about 1\,K, the anomaly smears out. Consequently
deviations from the ES law occur. Above about 6\,K, the anomaly has
vanished, and the DOS becomes constant. The sample is then in the
Mott regime. According to our interpretation, the cross-over between
ES and Mott law at such low temperatures is not due to the
temperature-enhanced range of hopping energies. But it origins from the
suppression of the Coulomb gap by thermal excitations. These
break-junction tunneling results are consistent with the electrical resistivity
of the same samples, which also shows a rather wide temperature range
for the transition between ES and Mott variable-range hopping.

\paragraph*{Acknowledgments:}
This work was supported by the German SFB 252
Darmstadt/Frankfurt/Mainz and the Russian RFFI 97-02-18280.


\begin{figure}
\caption{
$dI/dU$ vs.~$U$ spectra of the Ge sample with $N_d=1.26\cdot10^{17}\,{\rm{cm}}^{-3}$.
}
\label{fig-1}
\end{figure}

\begin{figure}
\caption{
Electrical resistance of the bulk Ge sample $R$ vs.~$T^{1/4}$ and $T^{1/2}$, respectively.
$N_d=1.26\cdot10^{17}\,{\rm{cm}}^{-3}$.
}
\label{resistance}
\end{figure}

\begin{figure}
\caption{
Coulomb gap $\Delta_{\rm{CG}}$ and residual DOS $\gamma$ vs.~ temperature
$T$. $N_d=1.26\cdot10^{17}\,{\rm{cm}}^{-3}$.
}
\label{fig-3}
\end{figure}


\end{document}